\documentclass[preprint,prd,eqsecnum,aps,epsf]{revtex4}

\begin{document}

\def\aprge{\buildrel > \over {_{\sim}}}
\def\aprle{\buildrel < \over {_{\sim}}}

\def\etal{{\it et.~al.}}
\def\ie{{\it i.e.}}
\def\eg{{\it e.g.}}

\def\bwt{\begin{widetext}}
\def\ewt{\end{widetext}}
\def\be{\begin{equation}}
\def\ee{\end{equation}}
\def\bea{\begin{eqnarray}}
\def\eea{\end{eqnarray}}
\def\bean{\begin{eqnarray*}}
\def\eean{\end{eqnarray*}}
\def\bary{\begin{array}}
\def\eary{\end{array}}
\def\bi{\bibitem}
\def\bit{\begin{itemize}}
\def\eit{\end{itemize}}

\def\lan{\langle}
\def\ran{\rangle}
\def\lra{\leftrightarrow}
\def\la{\leftarrow}
\def\ra{\rightarrow}
\def\dash{\mbox{-}}
\def\ol{\overline}

\def\ub{\ol{u}}
\def\db{\ol{d}}
\def\sb{\ol{s}}
\def\cb{\ol{c}}

\def\re{\rm Re}
\def\im{\rm Im}

\def \b{{\cal B}}
\def \ca{{\cal A}}
\def \ko{K^0}
\def \ok{\overline{K}^0}
\def \s{\sqrt{2}}
\def \st{\sqrt{3}}
\def \sx{\sqrt{6}}
%\begin{document}
%\begin{large}
\title{{\bf Anomalous Action of QCD from the General Quark Propagator}}
\author{Yong-Liang Ma$^\dag$, Qing Wang$^\ddag$  and Yue-Liang Wu$^\dag$}
\address{$^\dag$Institute of Theoretical Physics, Chinese Academy of Sciences, Beijing
100080, China\\
$^\ddag$Department of Physics, Tsinghua University, Beijing
100084, China}
\date{\today}
%%%%%%%%%%%%%%%%%%%%%%%%%%%%%%%
\begin{abstract}

The anomalous action of the chiral effective theory to
$\mathcal{O}(p^4)$ is investigated by generalizing the
consideration in \cite{ma02} with including the wave function part
in the general quark propagator. It is found that the QCD dynamics
dependence of the Wess-Zumino term is explicit and this dependence
means that the QCD chiral symmetry must be broken dynamically in
the low energy region for dynamically generating the Wess-Zumino
term. In addition, we found that the next to leading order
anomalous action in the momentum expansion of the chiral
perturbation theory is gauge invariant and QCD dynamics dependent.
\end{abstract}
\maketitle
%%%%%%%%%%%%%%%%%%%%%%%%%%%%%%%%%%%%%%%%%%%%%%%%%%%%%%%%%%%%%%%%%%%%%%%%%%%%%%

\section{introduction}

\label{sec Int}

Due to the quark confinement in the low energy region, the strong
interaction processes cannot be calculated in the frame work of
conventional perturbation theory in which the expansion is in
terms of the coupling constant $g_s$. In this sense, to describe
the low energy dynamics of QCD, some effective theories, based on
the Weinberg's folk theorem\cite{folk-theorem} that any quantum
theory that at sufficiently low energy and large distances looks
Lorentz invariant and satisfied the cluster decomposition
principle will also at sufficiently low energy look like a quantum
field theory, was proposed. Among all these effective theories,
chiral perturbation theory\cite{folk-theorem,Gasser} got great
triumphes in describing the hadron processes with pseudoscalar
mesons.

In principle, there are two ways to obtain the chiral perturbation
theory. One is the phenomenological
construction\cite{folk-theorem,Gasser}, in which, chiral symmetry
$SU(3)_R\times SU(3)_L$ and its breaking $SU(3)_R\times
SU(3)_L\rightarrow SU(3)_V$ of QCD was considered, the
pseudoscalar mesons emerge from the chiral symmetry breaking as
Goldstone bosons and live in the coset space of subgroup of
original chiral symmetry. The lagrangian is expressed in terms of
the order of small quantities such as the ratio of external
momenta of mesons to the chiral symmetry breaking scale, and in
front of each independent structure term, a coefficient is given.
It is believed that the QCD dynamics are incorporated in the
unknown coefficients and these coefficients can be determined from
the underlying QCD in principle, but in practice they are
determined phenomenologically due to our few knowledge about the
nonperturbative calculation. In this approach, the pseudoscalar
mesons get mass through the Gell-Mann, Oakes and Renner
relations\cite{Gell-Mann} by adding the quark masses to the chiral
effective theory which breaks the chiral symmetry explicitly. In
fact, this method had been used to describe the hadron physics
before the appearance of QCD\cite{beforeQCD}.

Another approach is to obtain the chiral perturbation theory from
QCD\cite{WangQ1}, in which, the relation between the
phenomenological lagrangian and QCD was established. It is found
that the coefficients of the phenomenological lagrangian can be
expressed in terms of the Green functions of quarks so that they
can be regarded as the fundamental QCD definitions of the
coefficients of the phenomenological lagrangian. Then, unlike the
phenomenological construction in which the coefficients were
determined phenomenologically, in the QCD approach these
coefficients can be evaluated numerically from the first principle
of QCD as long as we have found method to compute all related
Green's functions of QCD. As the first approximation, the explicit
results were given in \cite{H.Yang}, which can be shown further to
be equivalent to the computation based on a gauge invariant,
nonlocal, dynamical model (GND model) \cite{H.Yang1}, in which all
the coefficients of the chiral effective lagrangian were expressed
in terms of the quark self-energy and the numerical results based
on the quark self-energy determined in some models were also
given, these numerical results coincide with the experimental
ones. In fact, there was another approach by using the anomaly
method\cite{anom-norm}, but the signs of the numerical results of
coefficients $L_7$ and $L_8$ predicted in this method are
different from the phenomenological results. In addition, it seems
that all the coefficients in this anomaly approach is independent
of the QCD dynamics. In fact, it is shown\cite{H.Yang} that the
anomalous contributions are cancelled by some parts of the normal
contributions and the cancellation leaves the quark self-energy
dependent contributions to the coefficients of the chiral
lagrangian.

What we mentioned above are mainly the normal part of the chiral
lagrangian. Besides the normal part, there is anomalous part with
odd intrinsic parity of the chiral lagrangian which is due to the
negative parity of the pseudoscalar mesons. The anomalous part of
the chiral lagrangian was first given in\cite{wess} by integrating
the consistent condition of anomaly in the language of current
algebra. Latter, this lagrangian was constructed
geometrically\cite{witten} with photon as an external field. Its
topological meaning was studied in detail by several
groups\cite{topological,C-W}. Like the normal part of the chiral
effective theory, the quantum corrections do not renormalize the
coefficient of the Wess-Zumino term since it is the leading order
of the anomalous action. For the loop corrections, it is found
that they are gauge invariant and give rise the higher order
anomaly\cite{prove-p6} since the Wess-Zumino term is not gauge
invariant, the explicit form of the next to leading order were
give in\cite{p6-form}. This does not mean that the
nonrenormalization theorem is violated, because only the summation
of all the orders of the effective theory is equivalent to the
underlying theory. For the anomalous part of the chiral effective
theory, as for the normal part, among all these constructions, the
QCD effects were hidden in some constants such as the pion decay
constants, so that the explicit effects of the QCD dynamics were
also not clear. Because of all these reasons, it is interesting to
investigate the explicit QCD dynamics dependence of the anomalous
action. In our previous work\cite{ma02}, in the fame work of GND
model, we found that the anomalous terms come from both dynamics
dependent and independent sources, after some cancellations, the
Wess-Zumino term is yielded and it is found that the coefficient
relates to the quark self-energy. In addition, when the quark
self-energy vanishes, the anomalous action also vanishes, this
means that the anomalous action is a QCD dynamics dependent
quantity. In fact, before our work, the dependence of strong
dynamics of Wess-Zumino term was investigated in the constituent
quark model by introducing a constituent quark masses
$M_Q$\cite{anom-constituent} and it is found that the Wess-Zumino
term vanishes when the strong interaction was switched off by set
$M_Q=0$. But in all their approximations, the hard constituent
quark masses $M_Q$ will cause wrong bad ultraviolet behavior.  The
anomalous section of the effective action and its applications
were reviewed in\cite{anom-review}.

In fact, the quark propagator in GND quark model is first order
approximation of the dynamical perturbation theory\cite{DPT}, and
when one switches off the external sources, only the quark
self-energy effect is considered. Generally speaking, the inverse
of quark propagator takes the form
\begin{eqnarray}
S^{-1}(k)=iA(k^2)k\hspace{-0.2cm}\slash-B(k^2)\label{quarkP}
\end{eqnarray}
where $B(k^2)$ arises from quark self-energy and the quark
condensation which indicates the chiral symmetry breaking if
\begin{eqnarray}
\{S^{-1}(p),\gamma_5\}=-2\gamma_5B(p^2)\neq0
\end{eqnarray}
and $A(k^2)$ is quark wave function part. In the high energy
region, their explicit forms can be determined order by order
explicitly through the ordinary perturbation theory, but we cannot
get their analytic forms without approximation in the infrared
region yet since we do not know how to deal with nonperturbation
theory.

As we mentioned before, in leading order of dynamical perturbation
theory $A(k^2)=1$, in addition, Schwinger-Dyson equation tells us
that $A(k^2)=1$ only happens in Landau gauge with bare gluon-quark
vertex and bare gluon propagator in the kernel of the integration
equation. So that, to respect real original QCD effects, we have
to include the $A(k^2)$ effect. In the previous work \cite{ma02}
and \cite{H.Yang1}, it is for simplicity of computation, we take
the approximation $A(k^2)=1$. The main purpose of present paper is
to overcome this shortcoming and investigate the effect of
$A(k^2)$.

In literatures, $Z(k^2)=1/A(k^2)$ and $\Sigma(k^2)=B(k^2)/A(k^2)$
are called the quark dressing function and mass function
respectively\cite{lattice1}. In the language of renormalization,
the quark propagator (\ref{quarkP}) depends on the renormalization
point $\mu$, i.e.
\begin{eqnarray}
S^{-1}(\mu,k)=iA(\mu,k^2)k\hspace{-0.2cm}\slash-B(\mu,k^2)
\end{eqnarray}
and in the standard momentum subtraction scheme
\begin{eqnarray}
&&Z(\mu,\mu^2)=1\\
&&\Sigma(\mu^2)=B(\mu,\mu^2)/A(\mu,\mu^2)=m(\mu)
\end{eqnarray}
And the scale $\mu$ characterized the mode of chiral symmetry
breaking. As was shown in\cite{pole1}, when the scale is large
enough which makes QCD in the perturbation region, quantum
corrections cannot break the chiral symmetry dynamically if the
current quark mass which breaks the chiral symmetry explicitly was
not introduced. While in the low energy, the chiral symmetry is
broken due to the quark condensation and the mixed quark-gluon
condensation. In case that the current quark mass is introduced,
the chiral symmetry will be broken explicitly due to the small
light quark masses when scale $\mu$ is large, but there is no
Wess-Zumino anomaly generated. When $\mu$ becomes so small that
the theory is in the nonperturbative region, the dynamical chiral
symmetry breaking term becomes larger than current quark mass and
the the chiral symmetry is broken dynamically. Considering this,
at high energy region, the renormalized quark propagator can be
related to the bare one through the wave function renormalization
constant $Z_2$ through
\begin{eqnarray}
S^{bare}(c,k)=Z_2(\mu,c)S(\mu,k)
\end{eqnarray}
where $c$ is a small constant which is induced in a regularization
such as $\epsilon$ in dimensional regularization and describes the
pole behavior of $Z_2$. Since the physical or the renormalized
propagator $S(\mu,k)$ is finite, the bare propagator should be
negative power of $c$ which is needed to cancel the infinities
arise from the loop integral. Then, for a sufficiently small $c$,
the relations between the wave function renormalization $Z_2$ and
the quark dressing function $Z$ at different renormalization point
can be yielded as
\begin{eqnarray}
&&\frac{Z_2(\mu_1,c)}{Z_2(\mu_2,c)}=\frac{Z(\mu_2,c)}{Z(\mu_1,c)}\\
&&\Sigma(k^2)=\Sigma(\mu_1,k^2)=\Sigma(\mu_2,k^2)
\end{eqnarray}
which means that mass function must be renormalization point
independent and a change of the renormalization point is just an
overall rescaling of $Z(\mu,k^2)$ by a momentum independent
constant. Then, the measure of the nonperturbative physics is the
deviation of $Z(\mu, k^2)$ from $1$ and the difference of
$\Sigma(k^2)$ from the renormalizaed quark mass
$m(\mu)$\cite{lattice2}. In this paper we will not distinguish the
concepts of quark dressing function and the wave function
renormalization constant and call $A(k^2)$ as the wave function
renormalization constant.

Formally, the functions $A(k^2)$ and $B(k^2)$ are constrained by
the Schwinger-Dyson equation. Although Schwinger-Dyson equation
may give a strong constraint on the dress function $A(k^2)$ and
quark self-energy $B(k^2)$, it cannot be solved exactly since it
is a recurrent integral equation. Based on some approximations,
special forms of $A(k^2)$ and $B(k^2)$ were got from the
Schwinger-Dyson equation in Landau gauge\cite{DPT,Landau}, axial
gauge\cite{axial} and covariant gauge\cite{covariant}.

In this paper, we will study the anomalous action based on the
general quark propagator (\ref{quarkP}) without discussing the
implications of the dynamical chiral symmetry breaking which have
been discussed in\cite{CSB} and also the amplitude of the scale
$\mu$, we only consider the theory in the nonperturbative region.
For the convenience of our discussion, we will introduce the
vector, axial-vector, scalar and pseudoscalar sources like what
was done in the normal section of the chiral effective
lagrangian\cite{Gasser}. The anomalous action we will investigate
is the part of the action proportional to odd number of
Levi-Civita tensor
$\epsilon_{\mu\nu\alpha\beta}$\cite{anom-review,ball}.

This paper was organized as follows: In section \ref{sec Var}, we
will discuss the meaning of chiral symmetry by considering the
relation between the chiral effective lagrangian and the effective
action we are interested in and give chiral invariant general
quark propagator. In section \ref{sec Cal}, we shall calculate the
Wess-Zumino term from the general quark propagator in section
\ref{sec Var} and investigate the effects of QCD dynamics in the
anomalous section of the chiral effective lagrangian. Our
conclusions and outlooks are presented in Section \ref{sec
Conclu}.

\section{Variations of the external sources under chiral transformation.}

\label{sec Var}

In this section, we shall analyze some transformation properties
of the external sources under chiral rotation that we will use in
the following calculations. The explicit action which is
proportional to the inverse of full propagator (\ref{quarkP}) is
presented when we switch off the external sources by comparing the
Gasser-Leutwyler lagrangian\cite{Gasser}.

It is found that the introduction of some external sources in the
study of the chiral perturbation is convenient. Following
ref.\cite{Gasser}, in four dimensional Euclidean space-time, we
consider pure quark kinetic part of bare QCD action in the
presence of external scalar, pseudoscalar, vector and axial-vector
sources
\begin{eqnarray}
S[\psi,\bar{\psi},J]&=&\int d^4x\bar{\psi}D\psi\nonumber\\
&=&\int
d^4x[\bar{\psi}\partial\hspace{-0.2cm}\slash\psi+\bar{\psi}J(x)\psi]\label{acion}
\end{eqnarray}
where the external sources are defined as
\begin{eqnarray}
J(x)=-iv\hspace{-0.2cm}\slash(x)-ia\hspace{-0.2cm}\slash(x)\gamma_5-s(x)+ip(x)\gamma_5
\end{eqnarray}
and quark mass matrix can be extracted from the scalar external
source $s$. In chiral perturbation theory, the power counting of
the vector $v_\mu$ and axial-vector $a_\mu $ sources are both
$\mathcal{O}(p)$ while scalar $s$ and pseudoscalar $p$ sources are
both $\mathcal{O}(p^2)$. For the following convenience, we rewrite
the Dirac operator $D$ as
\begin{eqnarray}
D\equiv
\nabla\hspace{-0.3cm}\slash-s+ip\gamma_5,~~~~~~\nabla\hspace{-0.3cm}\slash=\gamma^\mu\nabla_\mu~~~~~~\nabla_\mu\equiv\partial_\mu-iv_\mu-ia_\mu\gamma_5=-\nabla^\dag_\mu
\end{eqnarray}

Formally, the generating functional of QCD is
\begin{eqnarray}
Z[J]=\int\mathcal{D}\psi\mathcal{D}\bar{\psi}\mathcal{D}\Psi\mathcal{D}\bar{\Psi}\mathcal{D}A_\mu
e^{\int
d^4x[\mathcal{L}(\psi,\bar{\psi},\Psi,\bar{\Psi},A_\mu)+\bar{\psi}J\psi]}
\end{eqnarray}
where $\mathcal{L}(\psi,\bar{\psi},\Psi,\bar{\Psi},A_\mu)$ is
lagrangian of QCD with $\psi$, $\Psi$ and $A_\mu$ are light, heavy
quarks and gluon fields respectively. By integrating out the quark
and gluon fields and integrating in the pseudoscalar mesons, one
can formally rewrite the QCD generating functional as
\begin{eqnarray}
Z[J]=\int\mathcal{D}U e^{S_{\rm GL}[U,J]}\label{functionalmeson}
\end{eqnarray}
where $S_{\rm GL}[U,J]$ is the chiral effective lagrangian of
pseudoscalar mesons with external sources. In principle, $S_{\rm
GL}[U,J]$ consists normal part and anomalous part due to the
intrinsic parity of pseudoscalar mesons, that is
\begin{eqnarray}
S_{\rm GL}[U,J]=S_{\rm normal}[U,J]+S_{\rm anomaly}[U,J]
\end{eqnarray}
the normal part $S_{\rm normal}[U,J]$ was given in\cite{Gasser} to
$\mathcal{O}(p^4)$ and the leading anomalous part $S_{\rm
anomaly}[U,J]$ can be found in\cite{wess,witten} and next to
leading part was given in\cite{p6-form}.

Then, the action (\ref{acion}) which we are interested relates the
generating functional through
\begin{eqnarray}
Z[J]=\int\mathcal{D}U\mathcal{D}\psi\mathcal{D}\bar{\psi}
e^{S_{\rm eff}[\psi,\bar{\psi},U,J]}\label{functionalnonp}
\end{eqnarray}

By comparing (\ref{functionalmeson}) and (\ref{functionalnonp}),
we yield the relation between the chiral effective action
(\ref{functionalmeson}) and the effective action
(\ref{functionalnonp})
\begin{eqnarray}
e^{S_{\rm GL}[U,J]}=\int\mathcal{D}\psi\mathcal{D}\bar{\psi}
e^{S_{\rm eff}[\psi,\bar{\psi},U,J]}\label{GL-eff}
\end{eqnarray}

Because the massless QCD is invariant under chiral transformation,
the action $S_{\rm eff}[\psi,\bar{\psi},U,J]$ is required to be
invariant under the following local $SU(3)_R\times SU(3)_L$ chiral
transformation
\begin{eqnarray}
\psi(x)&\rightarrow&\psi^\prime(x)=[V_L(x)P_L+V_R(x)P_R]\psi(x)\\
J(x)&\rightarrow&J^\prime(x)=[V_L(x)P_R+V_R(x)P_L][J(x)+\partial\hspace{-0.2cm}\slash_x][V_L^\dag(x)P_L+V_R^\dag(x)P_R]\\
U(x)&\rightarrow&U^\prime(x)=V_R(x)U(x)V_L^\dag(x)
\end{eqnarray}
where $V_L(x)$ and $V_R(x)$ are the left and right chiral rotation
matrices and $P_{R,L}$ are the project operators
${1\over2}(1\pm\gamma_5)$.

Mathematically, the pseudoscalar field $U(x)$ has the
decomposition $U(x)=\Omega^2(x)$, under local chiral
transformation $SU(3)_R\times SU(3)_L$, $\Omega(x)$ transforms as
\begin{eqnarray}
\Omega(x)\rightarrow\Omega^\prime(x)=h^\dag(x)\Omega(x)V_L^\dag(x)=V_R(x)\Omega(x)h(x)
\end{eqnarray}
and the gauge group $h(x)$ can be determined by $V_L$, $V_R$ and
$\Omega(x)$. The symmetry $h(x)$ is the so called hidden local
symmetry, its gauge bosons can be identified with the vector
mesons through the so called hidden local symmetry
method\cite{hidden,MWW-hidden}.

In terms of the nonlinear field $\Omega(x)$, we can introduce the
rotated external fields
\begin{eqnarray}
J_\Omega(x)&=&[\Omega(x)P_R+\Omega^\dag(x)P_L][J(x)+\partial\hspace{-0.2cm}\slash_x][\Omega^\dag(x)P_L+\Omega(x)P_R]\nonumber\\
&\equiv&-iv\hspace{-0.2cm}\slash_\Omega(x)-ia\hspace{-0.2cm}\slash_\Omega(x)\gamma_5-s_\Omega(x)+ip_\Omega(x)\gamma_5\label{rotate-source}
\end{eqnarray}
Replacing all the external fields with the rotated ones and
defining
\begin{eqnarray}
D_\Omega=\nabla\hspace{-0.27cm}\slash_\Omega-s_\Omega+ip_\Omega\gamma_5,~~~~~~~~
\nabla^\mu_\Omega\equiv\partial^\mu-iv^\mu_\Omega-ia^\mu_\Omega\gamma_5
\end{eqnarray}
then, under chiral transformation $SU(3)_R\times SU(3)_L$, one can
easily prove the transformations
\begin{eqnarray}
J_\Omega(x)&\rightarrow&J_\Omega^\prime(x)=h^\dag(x)[J_\Omega(x)+\partial\hspace{-0.2cm}\slash_x]h(x)\nonumber\\
D_\Omega&\rightarrow&D_\Omega^\prime=h^\dag(x)D_\Omega h(x)
\end{eqnarray}
From above transformations, one can conclude that the local chiral
transformation can be realized through hidden local symmetry, once
the theory keeps the hidden local symmetry, it preserves the
chiral symmetry automatically. That is, by introducing the
$\Omega$ field, the chiral symmetry can be reflected through the
hidden local symmetry. In fact, even in the presence of massive
quark, by replacing the external fields with the rotated ones, the
local chiral symmetry can be preserved\cite{ESP}.

For the convenience in our discussion of the Wess-Zumino term, we
extend $\Omega(x)$ into five dimensional space\cite{extra-dim}
with
\begin{eqnarray}     %(4)
\Omega(t,x)=e^{it\lambda^a\pi^a(x)}~~~~~~\Omega(x)=\Omega(1,x)~~~~~~\Omega(0,x)=1
\end{eqnarray}
and define the $t$ dependent rotated sources
\begin{eqnarray}     %(5)
J_\Omega(t,x)&\equiv& -iv\!\!\! /\;_\Omega(t,x)
-ia\!\!\! /\;_\Omega(t,x)\gamma_5-s_\Omega(t,x)+ip_\Omega(t,x)\gamma_5\nonumber\\
&=&[\Omega(t,x)P_R+\Omega^{\dagger}(t,x)P_L]
~[J(x)+\partial\!\!\!\! /\;_x]
~[\Omega(t,x)P_R+\Omega^{\dagger}(t,x)P_L] \label{JOmegadef}
\end{eqnarray}
we can verify
\begin{eqnarray}
&&s_\Omega(t,x)={1\over2}[\Omega(t,x)[s(x)-ip(x)]\Omega(t,x)
+\Omega^\dag(t,x)[s(x)+ip(x)]\Omega^\dag(t,x)]\\
\label{chirals}
&&p_\Omega(t,x)={i\over2}[\Omega(t,x)[s(x)-ip(x)]\Omega(t,x)
-\Omega^\dag(t,x)[s(x)+ip(x)]\Omega^\dag(t,x)]\label{chiralp}\\
&&v^{\mu}_\Omega(t,x)={1\over2}[
\Omega^{\dagger}(t,x)[v_{\mu}(x)+a_{\mu}(x)+i\partial_{\mu}]\Omega(t,x)\nonumber\\
&&\;\;\;\;\;\;\;\;\;\;\;\;\;\;\;\;\;\;\;+\Omega(t,x)[v_{\mu}(x)-a_{\mu}(x)+i\partial_{\mu}]\Omega^{\dagger}(t,x)]
\label{chiralv}\\
&&a^{\mu}_\Omega(t,x)={1\over2}[
\Omega^{\dagger}(t,x)[v_{\mu}(x)+a_{\mu}(x)+i\partial_{\mu}]\Omega(t,x)\nonumber\\
&&\;\;\;\;\;\;\;\;\;\;\;\;\;\;\;\;\;\;\;-\Omega(t,x)[v_{\mu}(x)-a_{\mu}(x)+i\partial_{\mu}]\Omega^{\dagger}(t,x)]
\label{chirala}
\end{eqnarray}

Now consider infinitesimal transformation at parameter space
$t\rightarrow t+\delta t$, correspondingly
\begin{eqnarray}
\Omega(t,x)&\rightarrow&\Omega(t+\delta t,x)=(1+i\delta
t\lambda^a\pi^a(x))\Omega(t,x)=\Omega(t,x)(1+i\delta
t\lambda^a\pi^a(x))\\
\Omega^\dag(t,x)&\rightarrow&\Omega^\dag(t+\delta t,x)= (1-i\delta
t\lambda^a\pi^a(x)) \Omega^\dag(t,x)=\Omega^\dag(t,x)(1-i\delta
t\lambda^a\pi^a(x))
\end{eqnarray}
we can easily prove
\begin{eqnarray}
\delta J_\Omega(t,x)&=&\{i\delta
t\lambda^a\pi^a(x)\gamma_5,J_\Omega(t,x)+\partial\hspace{-0.2cm}\slash_x\}\\
\delta a^\mu_\Omega(t,x)&=&-{\delta t\over2}\bigg[\frac{\delta
U}{\delta
t}U^\dag,v^\mu_\Omega(t,x)]\label{deltA}\\
\delta\overline{\nabla}^{\mu}_\Omega(t,x)&=&{i\delta
t\over2}[\frac{\delta U}{\delta
t}U^\dag(t,x),a^{\mu}_\Omega(t,x)]\label{deltV}
\end{eqnarray}
And in addition
\begin{eqnarray}
\delta\overline{\nabla}_\Omega^2(t,x)&=&-\frac{\delta t}{2}\{
[\lambda^a\pi^a(x),a^{\mu}_\Omega(t,x)]\overline{\nabla}^{\mu}_\Omega(t,x)
+\overline{\nabla}^{\mu}_\Omega(t,x)[\lambda^a\pi^a(x),a^{\mu}_\Omega(t,x)]\}
\end{eqnarray}

For the dimensional extended Dirac operator $D_\Omega(t,x)$
\begin{eqnarray}
D_\Omega(t,x)\equiv \nabla\!\!\!\!
/\;^t_\Omega-s_\Omega(t,x)+ip_\Omega(t,x)\gamma_5=\partial\!\!\!\!
/\;+J_\Omega(t,x)
\end{eqnarray}
we have
\begin{eqnarray}
\delta D_\Omega(t,x)=\{i\delta
t\lambda^a\pi^a\gamma_5,D_\Omega(t,x)\}={1\over2}\{{\delta
U(t,x)\over\delta t }U^\dag(t,x)\gamma_5\delta t,D_\Omega(t,x)\}
\end{eqnarray}
Equivalently
\begin{eqnarray}
{\partial D_\Omega(t,x)\over\partial t}&=&{1\over2}\{{\partial
U(t,x)\over\partial t}U^\dag(t,x)\gamma_5,D_\Omega(t,x)\}\\
{\partial D^{\dag}_\Omega(t,x)\over\partial
t}&=&{1\over2}\{{\partial U^\dag(t,x)\over\partial
t}U(t,x)\gamma_5,D^{\dag}_\Omega(t,x)\}\nonumber\\
&=&-{1\over2}\{U^\dag(t,x){\partial U(t,x)\over\partial
t}\gamma_5,D^{\dag}_\Omega(t,x)\} \label{variD}
\end{eqnarray}

By now, the transformation properties of the external sources were
derived from the classical lagrangian (\ref{acion}), we want to
include the QCD dynamics effects next. These effects can be
calculated perturbatively at high energy region, however, at low
energy region, the quark-gluon coupling $g_s$ is strong and it is
not suitable to regard it as an expansion parameter. In this
sense, we do not know how to get the forms of the wave function
renormalization constant $A(k^2)$ and the quark self-energy
$B(k^2)$ although formally they satisfy the Schwinger-Dyson
equation. Since we have the conclusion that the Wess-Zumino term
is an intrinsic property of the non-Abelian gauge theory both
phenomenologically\cite{wess} and geometrically\cite{witten}, we
naively want to know whether we can get the Wess-Zumino term when
the QCD non-Abelian gauge interaction was included in the action
(\ref{quarkP}) and explore the contributions of the wave function
renormalization constant $A(k^2)$ and self-energy $B(k^2)$.

To exhibits the form of the action include QCD non-Abelian gauge
interaction and its dynamical effects, following the method given
in ref.\cite{H.Yang1}, let us turn to the relation (\ref{GL-eff})
between the Gasser-Leutwyler action $S_{\rm GL}$ and the effective
action $S_{\rm eff}$. By using the rotated basis, this relation
can be rewritten as
\begin{eqnarray}
e^{S_{\rm
GL}[U,J]}&=&\frac{\int\mathcal{D}\psi\mathcal{D}\bar{\psi}
e^{S_{\rm
eff}[\psi,\bar{\psi},U,J]}}{\int\mathcal{D}\psi\mathcal{D}\bar{\psi}
e^{S[\psi,\bar{\psi},J]}}\int\mathcal{D}\psi\mathcal{D}\bar{\psi}
e^{S[\psi,\bar{\psi},J]}\nonumber\\
&=&\mathcal{N}\frac{\int\mathcal{D}\psi_\Omega\mathcal{D}\bar{\psi}_\Omega
e^{S_{\rm
eff}[\psi_\Omega,\bar{\psi}_\Omega,1,J_\Omega]}}{\int\mathcal{D}\psi_\Omega\mathcal{D}\bar{\psi}_\Omega
e^{S[\psi_\Omega,\bar{\psi}_\Omega,J_\Omega]}}\label{rotated-GL-eff}
\end{eqnarray}
where $S[\psi,\bar{\psi},J]$ was defined by (\ref{acion}) and
$\mathcal{N}=det[\partial\hspace{-0.2cm}\slash+J]$. The anomaly
due to the variation of the integration measure was cancelled
between the denominator and the numerator. The relation
(\ref{rotated-GL-eff}) means that, formally, we can write $S_{\rm
eff}$ as
\begin{eqnarray}
S_{\rm
eff}[\psi_\Omega,\bar{\psi}_\Omega,1,J_\Omega]=S[\psi_\Omega,\bar{\psi}_\Omega,J_\Omega]+S_{\rm
int}[\psi_\Omega,\bar{\psi}_\Omega,J_\Omega]\label{S-eff}
\end{eqnarray}
where $S[\psi_\Omega,\bar{\psi}_\Omega,J_\Omega]$ is the action
defined in (\ref{acion}) by substituting relative quantities with
the rotated ones. $S_{\rm int}$ are the terms represent effects of
color gauge interaction which are proportional to $\alpha_s$ at
the leading order, it includes the fermion self-energy term and
the wave function renormalization constant which are caused by
integrating out the heavy fermion and gluon degree of freedom in
the underling QCD and integrating in the local pseudoscalar fields
$U(x)$. We formally write the interaction part $S_{\rm int}$ as
\begin{eqnarray}
S_{\rm int}[\psi_\Omega,\bar{\psi}_\Omega,J_\Omega]\sim\int
d^4x\bar{\psi}_\Omega(x)\{[A(\partial^2)-1](\partial\hspace{-0.2cm}\slash+\tilde{J}_\Omega)+B(\partial^2)\}\psi_\Omega(x)\label{action-int}
\end{eqnarray}
Where we have taken the minimal extension by including in effect
of quark wave function renormalization corrections into the
$S_{\rm int}$ and
\begin{eqnarray}
\tilde{J}_\Omega&\equiv& -iv\!\!\! /\;_\Omega(t,x) -ia\!\!\!
/\;_\Omega(t,x)\gamma_5
\end{eqnarray}

It should be noticed that (\ref{action-int}) is the most general
form when the dynamical effects are independent of external
sources. In the presence of external sources, the general
effective action (\ref{action-int}) due to the color gauge
interaction can be decomposed in terms of its general spinor
structure which will introduce many more other functions besides
the present $A(\partial^2)$ and $B(\partial^2)$, these extra
functions will cause much more computation difficulties and we
will leave the discussion of them in future. Now as first step
beyond original quark self-energy and reduce the difficulties of
the computation, we only consider one more extra function: quark
wave function renormalization function $A(\partial^2)$ but leaves
the discussion of most general propagator elsewhere. In
(\ref{action-int}), we select the dynamical dependence
coefficients of the differential operator and the vector source
are the same because we need the variety of vector sources to
compensate the variety induced by the differential operator under
the local chiral transformation, the coefficients of the vector
source and the axial-vector source are also the same as we
considered the combinations
\begin{eqnarray}
V_{R,\mu}^\Omega & = & {1\over2}[v_{\mu}^\Omega+a_\mu^\Omega]\\
V_{L,\mu}^\Omega & = & {1\over2}[v_{\mu}^\Omega-a_\mu^\Omega]
\end{eqnarray}
and the Parity invariance of QCD induced property
\begin{eqnarray}
P: V_{R,\mu}^\Omega \leftrightarrow V_{L,\mu}^\Omega
\end{eqnarray}

Combining this with (\ref{S-eff}), we yield the effective action
as
\begin{eqnarray}
S_{\rm eff}[\psi_\Omega,\bar{\psi}_\Omega,1,J_\Omega]\sim\int
d^4x~\bar{\psi}_\Omega(x)\{A(\partial^2)[\partial\hspace{-0.2cm}\slash+\tilde{J}_\Omega]+s_\Omega-ip_\Omega\gamma_5+B(\partial^2)\}\psi_\Omega(x)\label{action-effect}
\end{eqnarray}

Substituting (\ref{action-effect}) into (\ref{rotated-GL-eff}), we
get the full expression of $S_{\rm GL}$ as
\begin{eqnarray}
S_{\rm GL}[U,J]&\sim&\ln
\det\{A(\partial^2)[\partial\hspace{-0.2cm}\slash+\tilde{J}_\Omega]+s_\Omega-ip_\Omega\gamma_5+B(\partial^2)\}\nonumber\\
&&-\ln \det[\partial\hspace{-0.2cm}\slash+J_\Omega]+\ln
\det[\partial\hspace{-0.2cm}\slash+J]\label{action-GL}
\end{eqnarray}
It should be noticed that there is no factors $A(\partial^2)$ and
$B(\partial^2)$ in the last two determinants, this is because
these two factors originate from the color interaction and the
last two determinants are from the free quark lagrangian.

As we discussed before, the local chiral symmetry can be reflected
by the hidden local symmetry through the introduction of the
rotated sources, but it is obvious that the action
(\ref{action-GL}) cannot preserve the chiral symmetry because
\begin{eqnarray}
h^\dag(x)A(-\partial^2)h(x)=A[-h^\dag(x)\partial^2h(x)]=A[-[\partial_\mu+h^\dag(x)\partial_\mu
h(x)]^2]
\end{eqnarray}
Similarly for $B(-\partial^2)$. This means that the chiral
transformation always induces an extra term in $A(-\partial^2)$
which makes the theory change under chiral transformation. To
compensate this extra term we consider the operator
$A(-\bar{\nabla}_\Omega^2)$ in stead of $A(-\partial^2)$ in the
action with $\bar{\nabla}^\mu_\Omega=\partial^\mu-iv^\mu_\Omega$.
Because, under local chiral transformation
\begin{eqnarray}
\bar{\nabla}^\mu_\Omega\rightarrow\bar{\nabla}^{\prime\mu}_\Omega=h^\dag(x)\bar{\nabla}^\mu_\Omega
h(x)
\end{eqnarray}
that is, the variation of the vector source compensates the
variation induce by the action of the differential operator on the
local phase factor, we have
\begin{eqnarray}
A(-\bar{\nabla}_\Omega^2)\rightarrow
A(-\bar{\nabla}_\Omega^{\prime2})=h^\dag(x)A(-\bar{\nabla}_\Omega^2)h(x)
\end{eqnarray}
which indicates that $A(-\bar{\nabla}_\Omega^2)$ transforms
homogenously under chiral transformation. Using the same analysis
on $B(-\partial^2)$, we finally get the local chiral invariant
action as
\begin{eqnarray}
S_{\rm GL}[U,J]&\sim&\ln
\det\{A(-\bar{\nabla}_\Omega^2)[\partial\hspace{-0.2cm}\slash+\tilde{J}_\Omega]+s_\Omega-ip_\Omega\gamma_5+B(-\bar{\nabla}_\Omega^2)\}\nonumber\\
&&-\ln \det[\partial\hspace{-0.2cm}\slash+J_\Omega]+\ln
\det[\partial\hspace{-0.2cm}\slash+J]\label{ext-action-GL}
\end{eqnarray}
Now let's make some comments on this action: When we do not
consider the wave function renormalization, i.e., set $A=1$, we
arrive at the GND quark model and its anomaly had been analyzed
in\cite{ma02}. If the QCD dynamical effects were switched off,
that is, $A=1$ and $B=0$, the first two terms of this action
cancelled each other and the left term is independent of the
pseudoscalar fields, then one cannot yield the Wess-Zumino term in
this case. As shown in\cite{ma02}, the scalar source $s$ and
pseudoscalar source $p$ do not contribute to the leading order
anomaly, so that we will not consider these two sources in the
following, therefore, our final action can be written as
\begin{eqnarray}
S_{\rm GL}[U,\tilde{J}]&=&\ln \det[D^\prime_\Omega]-\ln
\det[\partial\hspace{-0.2cm}\slash+\tilde{J}_\Omega]+\ln
\det[\partial\hspace{-0.2cm}\slash+\tilde{J}]\label{EGND-action}
\end{eqnarray}
with
\begin{eqnarray}
D^\prime_\Omega&=&A(-\bar{\nabla}_\Omega^2)\nabla\hspace{-0.27cm}\slash_\Omega-B(-\bar{\nabla}_\Omega^2)
\end{eqnarray}

\section{calculation of anomalous effective lagrangian with the generalized fermion propagator}

\label{sec Cal}

In this section, based on the action (\ref{EGND-action}), we will
calculate the anomalous section of chiral effective action to the
leading order, $\mathcal{O}(p^4)$.  Since the difference of
(\ref{EGND-action}) with that in our previous work \cite{ma02} is
quark wave function renormalization function $A$, the questions
are interested become that: is there any changes due to arbitrary
$A$ for the leading order anomalous action?  Does this anomalous
action has its QCD dynamics dependence?  Is there higher order
anomalous action and where is it from? What is the meaning of
higher order anomalous action?

Our consideration is: If there is anomalous action, it should be
evaluated through the direct calculation of the action
(\ref{EGND-action}). To investigate the QCD dynamics effects in
the anomalous action, one can set $A(k^2)=1$ and $B(k^2)=0$ in the
action (\ref{EGND-action}) which mean that the strong interaction
effects were switched off. In addition, if there is anomalous
action by the total calculation of (\ref{EGND-action}), this
anomalous action can be evaluated through the terms in the r.h.s.
of (\ref{EGND-action}), so we need to calculate the anomalous
action from every term in the r.h.s.of (\ref{EGND-action}). Since
the first term in the r.h.s.of (\ref{EGND-action}) is QCD dynamics
dependent, we expect the exact form of the Wess-Zumino term and
the Bardeen anomaly will impose some constraints on the behavior
of the strong dynamics dependent function $A(k^2)$ and $B(k^2)$
even though this constraint may be rude.

\subsection{Direct Calculation of Anomalous Action Based on the Effective Action (\ref{EGND-action}).}

\label{ssec Total}

To investigate the anomalous action in the effective action
(\ref{EGND-action}), we make an inverse rotation of
(\ref{rotate-source}) which makes the last two terms of
(\ref{EGND-action}) cancel each other and the pseudoscalar fields
in the second term rotate to the first term, that is
\begin{eqnarray}
S_{\rm GL}[U,\tilde{J}]&=&\ln
\det\{\tilde{A}(-\bar{\nabla}_\Omega^2)\nabla\hspace{-0.27cm}\slash-\hat{B}(-\bar{\nabla}_\Omega^2)\}
\end{eqnarray}
where
\begin{eqnarray}     %(4)
\tilde{A}(-\bar{\nabla}_\Omega^2)&=&[\Omega^\dag(t,x)P_R+\Omega(t,x)P_L]A(-\bar{\nabla}_\Omega^2)[\Omega(t,x)P_R+\Omega^\dag(t,x)P_L]\\
\hat{B}(-\bar{\nabla}_\Omega^2)&=&[\Omega^\dag(t,x)P_R+\Omega(t,x)P_L]B(-\bar{\nabla}_\Omega^2)[\Omega^\dag(t,x)P_R+\Omega(t,x)P_L]
\end{eqnarray}

Considering identities $\ln\det A={\rm Tr}\ln A$ and $\delta {\rm
Tr}\ln A={\rm Tr}\delta AA^{-1}$, we have
\begin{eqnarray}
S[U,\tilde{J}]-S[1,\tilde{J}]&=&\int_0^1dt {\rm
Tr}\bigg[\frac{\partial \hat{D}_\Omega^\prime(t)}{\partial
t}\hat{D}_\Omega^\prime(t)^{-1}\bigg]\label{action-toatl}
\end{eqnarray}
where the trace ${\rm Tr}$ is around the color, configure, flavor
and Lorentz space and
\begin{eqnarray}
\hat{D}_\Omega^\prime(t)=\tilde{A}(-\bar{\nabla}_\Omega^2)\nabla\hspace{-0.27cm}\slash-\hat{B}(-\bar{\nabla}_\Omega^2)
\end{eqnarray}
It should be noticed that we introduced a term $S[1,\tilde{J}]$ in
(\ref{action-toatl}) which would not change our conclusion on the
Wess-Zumino term since it is independent of the pseudoscalar
fields.

Since the anomalous effective lagrangian is proportional to the
odd number of Levi-Civita tensor
$\epsilon_{\mu\nu\alpha\beta}$\cite{ball}, i.e., odd number of
$\gamma_5$ in the trace of Lorents matrices, in the following we
only concentrate on this kind of terms. After the trace around
configure space, we have the anomalous action as
\begin{eqnarray}
S[U,\tilde{J}]-S[1,\tilde{J}]&=&\int_0^1dt \int
d^4x\int\frac{d^4k}{(2\pi)^4}{\rm tr}_{c,f,l}\bigg[\frac{\partial
\hat{D}_{\Omega,\partial\rightarrow
ik+\partial}^\prime(t)}{\partial
t}\hat{D}_{\Omega,\partial\rightarrow
ik+\partial}^\prime(t)^{-1}\bigg]_\epsilon
\end{eqnarray}
the subindex $\epsilon$ means we are only interested in the terms
consist odd number of Levi-Civita tensors and $\partial\rightarrow
ik+\partial$ indicates that after the trace on the configure
space, the differential operator $\partial$ should be replaced by
$ik+\partial$, that is
\begin{eqnarray}
\hat{D}^\prime_{\Omega,\partial\rightarrow
ik+\partial}=\tilde{A}[-(\bar{\nabla}_\Omega+ik)^2][\nabla\hspace{-0.27cm}\slash+ik\hspace{-0.2cm}\slash~]-\hat{B}_\Omega[-(\bar{\nabla}_\Omega+ik)^2]
\end{eqnarray}

The next problem we have to solve is how to deal with the function
$\hat{A}[-(\bar{\nabla}_\Omega+ik)^2]$ and
$\hat{B}[-(\bar{\nabla}_\Omega+ik)^2]$. To illustrate this
problem, let us focus on $\hat{A}[-(\bar{\nabla}_\Omega+ik)^2]$,
to deal with this function, one has to deal with function
$A[-(\bar{\nabla}_\Omega+ik)^2]$ at first. To make its power
counting explicit, we expand it around $k^2$, then using Taylor
expansion, we have
\begin{eqnarray}
A[-(ik+\bar{\nabla}_\Omega)^2]&=&A[k^2-2ik\cdot\bar{\nabla}_\Omega-\bar{\nabla}_\Omega^2]\nonumber\\
&=&A(k^2)+\sum_{n=1}^\infty A^{(n)}(k^2)\frac{(-1)^n}{n!}(2ik\cdot\bar{\nabla}_\Omega+\bar{\nabla}_\Omega^2)^n\nonumber\\
&\equiv&A(k^2)+\mathcal{A}(k_\mu,\bar{\nabla}_\Omega^\mu)
\end{eqnarray}
where $\mathcal{A}$ is a function of external source and its
chiral counting is equal or higher than $\mathcal{O}(p)$,
explicitly, it is
\begin{eqnarray}
\mathcal{A}(k_\mu,\bar{\nabla}_\Omega^\mu)&=&\sum_{n=1}^\infty
A^{(n)}(k^2)\frac{(-1)^n}{n!}(2ik\cdot\bar{\nabla}_\Omega+\bar{\nabla}_\Omega^2)^n
\end{eqnarray}

So that
\begin{eqnarray}
\tilde{A}[-(\bar{\nabla}_\Omega+ik)^2]=A(k^2)+\tilde{\mathcal{A}}(k_\mu,\bar{\nabla}_\Omega^\mu)
\end{eqnarray}
with
\begin{eqnarray}
\tilde{\mathcal{A}}(k_\mu,\bar{\nabla}_\Omega^\mu)&=&[\Omega^\dag(t,x)P_R+\Omega(t,x)P_L]\mathcal{A}(k_\mu,\bar{\nabla}_\Omega^\mu)[\Omega(t,x)P_R+\Omega^\dag(t,x)P_L]
\end{eqnarray}
Similarly, we can write
\begin{eqnarray}
B[-(ik+\bar{\nabla}_\Omega)^2]=B(k^2)+\mathcal{B}(k_\mu,\bar{\nabla}_\Omega^\mu)
\end{eqnarray}
with
\begin{eqnarray}
\mathcal{B}(k_\mu,\bar{\nabla}_\Omega^\mu)&=&\sum_{n=1}^\infty
B^{(n)}(k^2)\frac{(-1)^n}{n!}(2ik\cdot\bar{\nabla}_\Omega+\bar{\nabla}_\Omega^2)^n
\end{eqnarray}
Then
\begin{eqnarray}
\hat{B}[-(\bar{\nabla}_\Omega+ik)^2]=\hat{B}(k^2)+\hat{\mathcal{B}}(k_\mu,\bar{\nabla}_\Omega^\mu)
\end{eqnarray}
with
\begin{eqnarray}
\hat{B}(k^2)&=&B(k^2)[U^\dag(t,x)P_R+U(t,x)P_L]\\
\hat{\mathcal{B}}(k_\mu,\bar{\nabla}_\Omega^\mu)&=&[\Omega^\dag(t,x)P_R+\Omega(t,x)P_L]\mathcal{B}(k_\mu,\bar{\nabla}_\Omega^\mu)[\Omega^\dag(t,x)P_R+\Omega(t,x)P_L]
\end{eqnarray}

Then, we can formally write
\begin{eqnarray}
\hat{D}^\prime_\Omega(t)&=&A(k^2)ik\hspace{-0.2cm}\slash-B(k^2)[U^\dag(t,x)P_R+U(t,x)P_L]+\mathcal{E}[\bar{\nabla}_\Omega^\mu,a_\Omega^\mu]
\end{eqnarray}
with
\begin{eqnarray}
\mathcal{E}[\bar{\nabla}_\Omega^\mu,a_\Omega^\mu]&=&\tilde{\mathcal{A}}(k_\mu,\bar{\nabla}_\Omega^\mu)ik\hspace{-0.2cm}\slash~+\{A(k^2)+\tilde{\mathcal{A}}(k_\mu,\bar{\nabla}_\Omega^\mu)\}\nabla\hspace{-0.27cm}\slash~-\hat{\mathcal{B}}(k_\mu,\bar{\nabla}_\Omega^\mu)
\end{eqnarray}

So that
\begin{eqnarray}
\hat{D}^{\prime}_\Omega(t)^{-1}&=&\bigg\{A(k^2)ik\hspace{-0.2cm}\slash-B(k^2)[U^\dag(t,x)P_R+U(t,x)P_L]+\mathcal{E}[\bar{\nabla}_\Omega^\mu,a_\mu]\bigg\}^{-1}\nonumber\\
&=&\{A(k^2)ik\hspace{-0.2cm}\slash-B(k^2)[U^\dag(t,x)P_R+U(t,x)P_L]\}^{-1}\nonumber\\
&&\times\{1+\mathcal{E}[k_\mu,
\bar{\nabla}_\Omega^\mu][A(k^2)ik\hspace{-0.2cm}\slash-B(k^2)[U^\dag(t,x)P_R+U(t,x)P_L]]^{-1}\}^{-1}\nonumber\\
&=&\frac{-A(k^2)ik\hspace{-0.2cm}\slash-B(k^2)[U(t,x)P_R+U^\dag(t,x)P_L]}{A^2(k^2)k^2+B^2(k^2)}\nonumber\\
&&\times\sum_{m=0}^\infty(-1)^m\{\mathcal{E}[\bar{\nabla}_\Omega^\mu,a_\Omega^\mu]\frac{-A(k^2)ik\hspace{-0.2cm}\slash-B(k^2)[U(t,x)P_R+U^\dag(t,x)P_L]}{A^2(k^2)k^2+B^2(k^2)}\}^{m}
\end{eqnarray}

By now, the calculation becomes straight forward although complex.
Explicit calculation gives
\begin{eqnarray}
S[U,\tilde{J}]-S[1,\tilde{J}]&=&-\frac{N_cC}{48\pi^2}\epsilon_{\mu\nu\alpha\beta}\int_0^1dt
\int d^4x\nonumber\\
&&\times{\rm tr}_{f}\bigg[U^\dag(t,x)\frac{\partial
U(t,x)}{\partial
t}L_\mu(t,x)L_\nu(t,x)L_\alpha(t,x)L_\beta(t,x)\bigg]\label{4-WZ}
\end{eqnarray}
where $N_c$ is the number of color,
$L_\mu(t,x)=U^\dag(t,x)\partial_\mu U(t,x)$ and
\begin{eqnarray}
C&=&\frac{3}{\pi^2}\int
d^4k\bigg[\frac{4A^4(k^2)B^6(k^2)}{[A^2(k^2)k^2+B^2(k^2)]^5}-\frac{8A^4(k^2)B^5(k^2)B^\prime(k^2)k^2}{[A^2(k^2)k^2+B^2(k^2)]^5}+\frac{8A^3(k^2)A^\prime(k^2)B^6(k^2)k^2}{[A^2(k^2)k^2+B^2(k^2)]^5}\bigg]\nonumber\\
&=&-12\int_0^\infty
dk^2\frac{A^2(k^2)B^4(k^2)k^2}{[A^2(k^2)k^2+B^2(k^2)]^3}\frac{d}{dk^2}\frac{B^2(k^2)}{[A^2(k^2)k^2+B^2(k^2)]}\label{def-C}
\end{eqnarray}
This constant shows the QCD dynamics dependence of the Wess-Zumino
term.

%%%%%%%%%%%%%%%%%%%%%%%%%%%%%%%%%%%%%%%%%%%%%%%%%%%%%%%%%%%%%%%%%%%%%%%%%%%%
%If the integral is exactly $1$, the conclusion (\ref{4-WZ}) is the        %
%exact Wess-Zumino term\cite{wess}.                                        %
%%%%%%%%%%%%%%%%%%%%%%%%%%%%%%%%%%%%%%%%%%%%%%%%%%%%%%%%%%%%%%%%%%%%%%%%%%%%

Comparing this conclusion with that given in\cite{ma02}, it is
seen that when we set $A(q^2)=1$ which means the leading order of
dynamical perturbation theory, the present result is the same as
that of\cite{ma02}. From the present expression (\ref{4-WZ}), it
is clear that the Wess-Zumino term arrives from the action
(\ref{EGND-action}) which depends on the dynamics of QCD, when the
dynamics of QCD is switched off by taking $A(k^2)=1$ and
$B(k^2)=0$, the coefficient $C$ also vanishes, this means that the
Wess-Zumino term is an intrinsic property of QCD and is a result
of the strong interaction. Note that the QCD dynamics dependence
of Wess-Zumino term has also been investigated in\cite{GCM} based
on the GCM model, but our result here has a factor
$k^2A(k^2)/[k^2A(k^2)+B^2(k^2)]$ difference with the one
in\cite{GCM}.

In the case of dynamical chiral symmetry breaking, i.e. $B(k^2)
\neq 0$, the pseudoscalar mesons emerge as the Goldstone bosons
and there will be anomalous processes which can be described by
the Wess-Zumino term\cite{wess}. In this case, one should have
$C=1$ to recover the exact coefficient of Wess-Zumino term. We
find that as long as the functions $A(k^2)$ and $B(k^2)$ behave as
\begin{eqnarray}
t\equiv\frac{A^2(k^2)k^2}{B^2(k^2)}\rightarrow\left\{\begin{array}{lcl}
\infty &~~~&k^2\rightarrow\infty \\ 0
&&k^2\rightarrow0\end{array}\right. \label{constraint-limit}
\end{eqnarray}
The coefficient $C$ can explicitly be integrated out
\begin{eqnarray}
C&=&-12\int_0^\infty
dt\frac{t}{[1+t]^3}\frac{d}{dt}\frac{1}{[1+t]}=1
\end{eqnarray}
which does coincides with the exact coefficient of Wess-Zumino
term.

Although relation (\ref{constraint-limit}) seems like a constraint
on the behavior of the function $B(k^2)/A(k^2)$, while it is very
crude. From the explicit calculations given in the literatures, we
can extract the ratio
\begin{eqnarray}
\frac{B(k^2)/A(k^2)}{k^2}&\rightarrow&\frac{4m_D^3}{k^4}\ln^\gamma(\frac{-k^2}{\mu^2}),~~~~\gamma=\frac{12}{33-2n}\label{QSE}\\
\frac{B(k^2)/A(k^2)}{k^2}&\rightarrow&\left\{%
\begin{array}{ll}
    \frac{[1-\frac{4}{\lambda}[1+\frac{3}{4}\frac{1}{\lambda C}]+\mathcal{O}(t^2)]^{1/2}}{t[-{1\over\lambda}+{2\over3}{1\over\lambda^2}t+\mathcal{O}(t^2)]}, & \hbox{$k^2\rightarrow0$;} \\
   \frac{[1-{1\over2}\lambda t^{-1}+\mathcal{O}(t^{-2})]^{1/2}}{t^2[-{1\over t}+{2\over\lambda}{1\over t^2}]}, & \hbox{$k^2\rightarrow\infty$.} \\
\end{array}%
\right.    \label{covariant}\\
\frac{B(k^2)/A(k^2)}{k^2}&\rightarrow&\left\{%
\begin{array}{ll}
    c_1\Phi(2,{5\over2};{-k^2\over\beta M^2})+c_2({-k^2\over\beta M^2})^{-3/2}\Phi({1\over2},{-1\over2};{-k^2\over\beta M^2}), & \hbox{$k^2\rightarrow0$;} \\
    \frac{c_1\Phi(2,{5\over2};{-k^2\over\beta M^2})+c_2({-k^2\over\beta M^2})^{-3/2}\Phi({1\over2},{-1\over2};{-k^2\over\beta M^2})}{k^2}, & \hbox{$k^2\rightarrow\infty$.} \\
\end{array}%
\right.    \label{axial}
\end{eqnarray}
where (\ref{QSE}) was given in\cite{DPT} by using the Landau
gauge, eq.(\ref{covariant}) is from the conclusion given
in\cite{covariant} based  on the covariant gauge and
eq.(\ref{axial}) was the result from the axial-gauge\cite{axial}.
In (\ref{QSE}), $n$ and $m_D$ are the number of flavors and
constant dynamical quark mass respectively. In (\ref{constraint}),
$\lambda$ is a constant due to the ghost self-energy at zero point
and $t\propto k^2$. $\Phi$ is the confluent hypergeometric
function, $\beta\propto \alpha_s$ and $M$ is the renormalization
point in (\ref{axial}). After some basic algebra, it is seen that
the above three results (\ref{QSE}-\ref{axial}) all satisfy the
condition eq.(\ref{constraint-limit}).

To rewrite the four dimensional Wess-Zumino term $(\ref{4-WZ})$ in
five dimensional space-time with four dimension space-time
boundary, we use the same trick as\cite{ma02}
\begin{eqnarray}
&&\frac{\partial}{\partial t}
{\rm tr}_f\{L_i(t,x)L_j(t,x)L_k(t,x)L_l(t,x)L_m(t,x)\}\epsilon^{ijklm}\nonumber\\
&&\;\;\;\;\;\;\;\;\;\;\;=5\frac{\partial}{\partial x^m}{\rm
tr}_f\{U^\dag(t,x)\frac{ U(t,x)}{\partial
t}L_i(t,x)L_j(t,x)L_k(t,x)L_l(t,x)\}\epsilon^{ijklm}
\end{eqnarray}
where $\epsilon^{ijklm}$ is a totally antisymmetric tensor. This
relation yields
\begin{eqnarray}
&&\int_Qd\Sigma^{ijklm}{\rm tr}_f\{L_i(1,x)L_j(1,x)L_k(1,x)L_l(1,x)L_m(1,x)\}\nonumber\\
&&\;\;\;\;\;\;\;\;\;\;=\int
d\Sigma^{ijklm}\int_0^1dt\frac{\partial}{\partial t}{\rm tr}_f\{L_i(t,x)L_j(t,x)L_k(t,x)L_l(t,x)L_m(t,x)\}\nonumber\\
&&\;\;\;\;\;\;\;\;\;\;=5\int
d^4x\int_0^1dt\epsilon^{\mu\nu\alpha\beta}{\rm
tr}_f\{U^\dag(t,x)\frac{\partial U(t,x)}{\partial
t}L_\mu(t,x)L_\nu(t,x)L_\alpha(t,x)L_\beta(t,x)\}
\end{eqnarray}
From this and using $C=1$ we can get the standard Wess-Zumino
term\cite{wess}
\begin{eqnarray}
\Gamma^-[U]&=&-\frac{N_c}{240\pi^2}\int_Qd\Sigma^{ijklm}{\rm
tr}_f\{L_i(1,x)L_j(1,x)L_k(1,x)L_l(1,x)L_m(1,x)\}\label{5-WZ}
\end{eqnarray}

In conclusion, it is seen that the Wess-Zumino term is an
intrinsic property of QCD and depends on the QCD dynamics. All the
QCD dynamics effects in the Wess-Zumino term can be collected into
the coefficient $C$ which was given by (\ref{def-C}) and, when the
coefficient approaches to unit, one arrives at the exact
Wess-Zumino term.  When one switches off the QCD dynamics, the
coefficient will vanish, which means a vanishing anomalous action
and no anomalous pseudoscalar processes occur.

\subsection{Anomalous Action from Each Term of the r.h.s. of Effective Action (\ref{EGND-action}).}

\label{ssec r.h.s.}

In above subsection, we have proved that there is leading order
anomaly in the action (\ref{EGND-action}), now let's investigate
the imaginary part of the action originates from every term of
(\ref{EGND-action}).

Let us look at the anomalous action from $\Gamma^-_{1}$ from the
first term of (\ref{EGND-action}). By using the same method as
that used in above subsection, explicit calculation shows that the
anomalous action $\Gamma^-_{1,D}$ due to the QCD dynamics of the
first term of (\ref{EGND-action}) is
\begin{eqnarray}
\Gamma^-_{1,D}&=&4iC_1\epsilon_{\mu\nu\alpha\beta}\int_0^1dt\int
d^4x\frac{\partial}{\partial t}{\rm
tr}_f\bigg[\overline{\nabla}_\mu\overline{\nabla}_\nu\overline{\nabla}_\alpha
a_\beta+\overline{\nabla}_\mu a_\nu a_\alpha
a_\beta\bigg]_\Omega\nonumber\\
&&+4iC_2\epsilon_{\mu\nu\alpha\beta}\int_0^1dt\int
d^4x\frac{\partial}{\partial t}{\rm
tr}_f\bigg[\overline{\nabla}_\mu\overline{\nabla}_\nu\overline{\nabla}_\alpha
a_\beta\bigg]_\Omega\nonumber\\
&&+4iC_3\epsilon_{\mu\nu\alpha\beta}\int_0^1dt\int
d^4x\frac{\partial}{\partial t}{\rm
tr}_f\bigg[\overline{\nabla}_\mu\overline{\nabla}_\nu\overline{\nabla}_\alpha
a_\beta-\overline{\nabla}_\mu a_\nu a_\alpha
a_\beta\bigg]_\Omega\label{anom-1}
\end{eqnarray}
where the index $\Omega$ means that all the external sources are
rotated sources and depend on the chiral angle $\Omega$. And the
definitions of the constants $C_i$ are defined as
\begin{eqnarray}
C_1&=&\int\frac{d^4k}{(2\pi)^4}\frac{A^2(k^2)B^4(k^2)}{[A^2(k^2)k^2+B^2(k^2)]^3}\frac{d}{dk^2}\frac{B^2(k^2)}{[A^2(k^2)k^2+B^2(k^2)]}\label{GInt-1}\\
C_2&=&\int\frac{d^4k}{(2\pi)^4}\frac{2A^4(k^2)B^2(k^2)k^2}{[A^2(k^2)k^2+B^2(k^2)]^3}\frac{d}{dk^2}\frac{B^2(k^2)}{[A^2(k^2)k^2+B^2(k^2)]}\label{GInt-2}\\
C_3&=&\int\frac{d^4k}{(2\pi)^4}\frac{A^6(k^2)k^4}{[A^2(k^2)k^2+B^2(k^2)]^3}\frac{d}{dk^2}\frac{B(k^2)}{[A^2(k^2)k^2+B^2(k^2)]}\label{GInt-3}
\end{eqnarray}

Using $(\ref{deltA})$ and $(\ref{deltV})$, we can prove relations
\begin{eqnarray}
\epsilon_{\mu\nu\alpha\beta}\frac{\partial}{\partial t}{\rm
tr}_f\bigg[
\overline{\nabla}_\mu\overline{\nabla}_\nu\overline{\nabla}_\alpha
a_\beta\bigg]_\Omega&=&{i\over2}\epsilon_{\mu\nu\alpha\beta}{\rm
tr}_f\{\frac{\partial U}{\partial
t}U^\dag[2a_\mu\overline{\nabla}_\nu\overline{\nabla}_\alpha
a_\beta+\overline{\nabla}_\mu\overline{\nabla}_\nu a_\alpha
a_\beta-a_\mu\overline{\nabla}_\nu
a_\alpha\overline{\nabla}_\beta\nonumber\\
&&-\overline{\nabla}_\mu a_\nu\overline{\nabla}_\alpha
a_\beta+a_\mu
a_\nu\overline{\nabla}_\alpha\overline{\nabla}_\beta+2\overline{\nabla}_\mu\overline{\nabla}_\nu\overline{\nabla}_\alpha\overline{\nabla}_\beta]_\Omega\}\label{variatthreedelt}\\
\epsilon_{\mu\nu\alpha\beta}\frac{\partial}{\partial t}{\rm
tr}_f\bigg[\overline{\nabla}_\mu a_\nu a_\alpha
a_\beta\bigg]_\Omega&=&{i\over2}\epsilon_{\mu\nu\alpha\beta}{\rm
tr}_f\{\frac{\partial U}{\partial t}U^\dag[2a_\mu a_\nu a_\alpha
a_\beta+2\overline{\nabla}_\mu a_\nu
a_\alpha\overline{\nabla}_\beta+a_\mu
a_\nu\overline{\nabla}_\alpha\overline{\nabla}_\beta\nonumber\\
&&-\overline{\nabla}_\mu a_\nu\overline{\nabla}_\alpha
a_\beta-a_\mu\overline{\nabla}_\nu
a_\alpha\overline{\nabla}_\beta+\overline{\nabla}_\mu\overline{\nabla}_\nu
a_\alpha a_\beta]_\Omega\}\label{variatthreea}
\end{eqnarray}
we get the anomalous action with form
\begin{eqnarray}
\Gamma^-_{1,D}&=&2N_cC_1\epsilon_{\mu\nu\alpha\beta}\int_0^1
dt\int d^4x {\rm tr}_f\bigg[ {\partial U\over\partial
t}U^\dag\nonumber\\
&&\times
\{2\bar{\nabla}_\mu\bar{\nabla}_\nu\bar{\nabla}_\alpha\bar{\nabla}_\beta+2a_\mu a_\nu\bar{\nabla}_\alpha\bar{\nabla}_\beta-2\bar{\nabla}_\mu a_\nu\bar{\nabla}_\alpha a_\beta+2\bar{\nabla}_\mu a_\nu a_\alpha\bar{\nabla}_\beta\nonumber\\
&&+2a_\mu\bar{\nabla}_\nu\bar{\nabla}_\alpha
a_\beta-2a_\mu\bar{\nabla}_\nu
a_\alpha\bar{\nabla}_\beta+2\bar{\nabla}_\mu\bar{\nabla}_\nu
a_\alpha a_\beta+2a_\mu a_\nu a_\alpha a_\beta\}_\Omega\bigg]\nonumber\\
&&+2N_cC_2\epsilon_{\mu\nu\alpha\beta}\int_0^1 dt\int d^4x {\rm
tr}_f\bigg[ {\partial U\over\partial
t}U^\dag\nonumber\\
&&\times\{4\bar{\nabla}_\mu\bar{\nabla}_\nu\bar{\nabla}_\alpha\bar{\nabla}_\beta+2a_\mu
a_\nu\bar{\nabla}_\alpha\bar{\nabla}_\beta-2\bar{\nabla}_\mu
a_\nu\bar{\nabla}_\alpha
a_\beta+4a_\mu\bar{\nabla}_\nu\bar{\nabla}_\alpha
a_\beta\nonumber\\
&&-2a_\mu\bar{\nabla}_\nu
a_\alpha\bar{\nabla}_\beta+2\bar{\nabla}_\mu\bar{\nabla}_\nu
a_\alpha a_\beta\}_\Omega\bigg]\nonumber\\
&&+2N_cC_3\epsilon_{\mu\nu\alpha\beta}\int_0^1 dt\int d^4x {\rm
tr}_f \bigg[{\partial U\over\partial
t}U^\dag\nonumber\\
&&\times\{2\bar{\nabla}_\mu\bar{\nabla}_\nu\bar{\nabla}_\alpha\bar{\nabla}_\beta-2\bar{\nabla}_\mu
a_\nu
a_\alpha\bar{\nabla}_\beta+2a_\mu\bar{\nabla}_\nu\bar{\nabla}_\alpha
a_\beta-2a_\mu a_\nu a_\alpha a_\beta\}_\Omega\bigg]\label{anom-2}
\end{eqnarray}

By using the behavior (\ref{constraint-limit}), we can exactly get
\begin{eqnarray}
C_1&=&C_2={1\over3}C_3={-1\over12}{1\over16\pi^2}\label{constraint}
\end{eqnarray}
With this equation, Anomalous action with external gauge fields
can be yielded
\begin{eqnarray}
\Gamma^-_{1,D}&=&
-{1\over32\pi^2}\varepsilon_{\mu\nu\alpha\beta}\int_0^1 dt{\rm
tr_f}\bigg[{\partial U\over\partial
t}U^\dag\nonumber\\
&&\times\bigg\{V_{\mu\nu}V_{\alpha\beta}+\{{2i\over3}a_\mu a_\nu,
V_{\alpha\beta}\}+{4\over3}d_\mu a_\nu d_\alpha
a_\beta+{8i\over3}a_\mu V_{\nu\alpha} a_\beta+{4\over3}a_\mu a_\nu
a_\alpha a_\beta\bigg\}_\Omega\bigg]\label{Bardeen}
\end{eqnarray}
where
\begin{eqnarray}
V_{\mu\nu}^\Omega&=&\partial_\mu v_\mu^\Omega-\partial_\nu
v_\mu^\Omega-i[v_\mu^\Omega,v_\nu^\Omega]\\
d_\mu a_\nu^\Omega&=&\partial_\mu
a_\nu^\Omega-i[v_\mu^\Omega,a_\nu^\Omega]
\end{eqnarray}
which has the same form as the gauge anomaly by replacing $U$ with
the gauge transformation $g$\cite{Bardeen}.

From the above relations such as (\ref{anom-1}) and
(\ref{anom-2}), we see that the quark dynamics plays an important
role in the form of anomalous action, if there is no quark
condensation, there is no anomalous action as mentioned
in\cite{ma02}. The role of quark wave function renormalization
function $A(k^2)$ is passive which play the key role only with
quark self energy $B(k^2)$ in the combination of
$\Sigma(k^2)=B(k^2)/A(k^2)$. In addition, the conclusion
(\ref{Bardeen}) means that the first part of (\ref{EGND-action})
contributes to the imaginary part of the action
(\ref{EGND-action}).

We now investigate how to get the Wess-Zumino term\cite{wess} from
the the anomalous action (\ref{anom-2}). The Wess-Zumino term is
the pure pseudoscalar mesons lagrangian, so that we should switch
off the external fields by setting $\tilde{J}=0$. In this case,
there are integrable conditions
\begin{eqnarray}
V_\Omega^{\mu\nu}&=&i[a_\Omega^\mu,a_\Omega^\nu]\label{inte-1}\\
d^\mu a_\Omega^\nu&=&d^\nu a_\Omega^\mu\label{inte-2}\\
a_\Omega^\mu&=&{i\over2}\Omega^\dag(t,x)[\partial^\mu
U(t,x)]\Omega^\dag(t,x)\label{inte-3}
\end{eqnarray}
Then, substitute the above three relations to the non-Abelian
anomaly $(\ref{Bardeen})$, we get
\begin{eqnarray}
\Gamma^-_{1,D}[U]&=&-\frac{N_c}{48\pi^2}\int_0^1dt\int
d^4x\epsilon_{\mu\nu\alpha\beta}\nonumber\\
&&\;\;\;\;\times{\rm tr}_f\{U^\dag(t,x)\frac{\partial
U(t,x)}{\partial t}L_\mu(t,x)L_\nu(t,x)L_\alpha(t,x)L_\beta(t,x)\}
\end{eqnarray}
where $L_\mu(t,x)\equiv U^\dag(t,x)\partial_\mu U(t,x)$.

By now, can we say that the leading order anomalous action of
(\ref{EGND-action}) comes from its first part? As argued
in\cite{wess}, the anomalous of term of the effective action is
not chiral invariant which seems conflict with the argument that
the first term of (\ref{EGND-action}) is chiral invariant, does
this means that our calculation is wrong? The answer is NO! This
is because, we only considered the imaginary part of the first
term of (\ref{EGND-action}) that depends on the QCD dynamical
effects, if one switch off these effects in the first term of
(\ref{EGND-action}) through $A(k^2)=1$ and $B(k^2)=0$, one can
easily find that there is another term which dependent on the
rotated sources which will give rise the anomaly obviously but
with a negative sign. Its contribution $\Gamma^-_{1,0}$ can be
calculated by using the Fujikawa's path integral method which is
shown below. Then, one totally has
\begin{eqnarray}
\Gamma^-_{1}=\Gamma^-_{1,0}+\Gamma^-_{1,D}
\end{eqnarray}

So that, totally, the leading order imaginary section of the
action induced by the QCD dynamical effects in the first term of
(\ref{EGND-action}) cancels that induced by the rotated sources
and leaves the next to leading order anomalous action, and, the
next to leading order anomalous action is chiral invariant.

Next, let us investigate the anomalous $\Gamma^-_{2}$ induced by
the last two terms of (\ref{EGND-action}). This anomaly can be
evaluated by the Fujikawa's method\cite{Fujikawa}.,i.e.,
\begin{eqnarray}
\Gamma^-_{2}&=&-{\rm
Tr}\ln[\partial\hspace{-0.25cm}\slash+\tilde{J}_\Omega]+{\rm
Tr}\ln[\partial\hspace{-0.25cm}\slash+\tilde{J}]\nonumber\\
&=&-\int_0^1dt{\rm
Tr}\{\frac{\partial \tilde{J}_{\Omega(t)}}{\partial
t}[\partial\hspace{-0.2cm}\slash+\tilde{J}_\Omega(t)]\}\nonumber\\
&=&-\lim_{\Lambda\rightarrow\infty}\int_0^1dt{\rm
Tr}\{\frac{\partial U}{\partial t}U^\dag\gamma_5
\exp[\frac{[\partial\hspace{-0.2cm}\slash+\tilde{J}_\Omega]^2}{\Lambda^2}]\}
\end{eqnarray}

After the trace on configure space, expanding the exponent, taking
the limit $\Lambda\rightarrow\infty$, keeping the
$\mathcal{O}(p^4)$ terms and the terms proportional to
$\epsilon_{\mu\nu\alpha\beta}$, we finally get the result
\begin{eqnarray}
\Gamma^-_2&=&-Tr\ln[\partial\hspace{-0.2cm}\slash+J_\Omega]+Tr\ln[\partial\hspace{-0.2cm}\slash+J]\nonumber\\
&=&-\frac{N_c}{32\pi^2}\epsilon_{\mu\nu\alpha\beta}\int_0^1dt\int
d^4x{\rm tr}_f\bigg\{\frac{\partial U}{\partial
t}U^\dag\nonumber\\
&&\times\bigg[V_{\mu\nu}V_{\alpha\beta}+{4\over3}d_\mu a_\nu
d_\alpha a_\beta+{2i\over3}\{V_{\mu\nu}, a_\alpha
a_\beta\}+{8i\over3}a_\mu V_{\alpha\beta}a_\nu+{4\over3}a_\mu
a_\nu a_\alpha a_\beta\bigg]_\Omega\bigg\}
\end{eqnarray}
which is the famous Bardeen anomaly\cite{Bardeen}. After using the
integrability condition (\ref{inte-1}-\ref{inte-3}), we can get
the Wess-Zumino term as mentioned above.

In summary, from the r.h.s. of (\ref{EGND-action}), the anomalous
action is
\begin{eqnarray}
\Gamma^-=\Gamma^-_{1,0}+\Gamma^-_{1,D}+\Gamma^-_{2}
\end{eqnarray}
with $\Gamma^-_{1,D}=\Gamma^-_{2}=-\Gamma^-_{1,0}$.

\section{Conclusions and outlooks}

\label{sec Conclu}

In conclusion, start from the general quark propagator, after
chiral rotation, we get the anomalous section of the chiral
effective action with external sources. For which we can yield the
Wess-Zumino term after switch off the external fields.

The anomalous action we obtained depends on the QCD dynamics
closely, when we switch off the QCD dynamics, the Wess-Zumino term
will vanish. That means that the anomalous processes described by
the Wess-Zumino term is a QCD processes. To investigate which term
the anomaly arise from, we explicitly calculated the anomaly
induced by each term of action (\ref{EGND-action}). We found that
the first term of (\ref{EGND-action}) can induce QCD dynamics
dependent and independent anomalous action but with a sign
difference then the leading order anomaly vanishes, this means
that the first term of (\ref{EGND-action}) only contribute to the
next to leading order anomaly and the next to leading anomaly is
chiral invariant. We also found that the last two terms of
(\ref{EGND-action}) also contribute to anomaly and with the same
sign of Wess-Zumino term when we switch off the external sources.

Then, similar to the conclusion given in\cite{ma02}, one observes
two kinds of alternative cancellations: One cancellation is
regarded as that the anomalous action arising from the QCD
dynamical independent section of the first term of
(\ref{EGND-action}) is cancelled by the anomaly due to the
dynamical dependent section of the first term of
(\ref{EGND-action}) and leaves the anomalous action from the
second term of (\ref{EGND-action}). In this cancellation, the
anomalous action is found to be QCD dynamics independent. An
alternative cancellation is that the anomalous induced by the QCD
dynamics independent section of the first term of
(\ref{EGND-action}) was cancelled by the second term of
(\ref{EGND-action}) which is also QCD dynamics independent. As a
consequence, one obtains the anomalous action from the QCD
dynamics dependent section of the first term of
(\ref{EGND-action}). In such a cancellation, the dependence of the
Wess-Zumino term on the QCD can be seen explicitly. By considering
the totally calculation given in the first subsection of section
\ref{sec Cal}, we believe that the second cancellation is more
reasonable.

The QCD dynamics dependence of the Wess-Zumino term reflects its
scale dependence. At low energy region, the dynamical mass is
dominate and chiral symmetry is broken dynamically. While at high
energy region, the chiral symmetry is explicitly broken due to the
small quark mass, and the chiral symmetry is preserved when the
quark mass is neglected, then, there will be no anomalous meson
process. This situation is very similar to the chiral anomaly
studied in ref.\cite{MW-chiral} based on the loop
regularization\cite{LR}, in which both the massless and massive
QCD will be anomaly free when the sliding energy scale $\mu_s$ is
large enough.

In our calculation, we dealt with the fermion determinant
explicitly by using the expansion of the QCD dependent functions
$A(k^2)$ and $B(k^2)$ around the external sources. Besides this
method, the anomalous action from the term depending on the QCD
dynamics can also be evaluated with Fujikawa's path integral
method\cite{Fujikawa}, the anomalous action based on the general
quark propagator can also be calculated with this method. But in
this method, the behavior of the quark self-energy and the wave
function renormalization constant should be determined under the
limit $\tau\rightarrow\infty$ with $\tau$ as the regulator
introduced in this method, this makes the effects of the dynamical
functions obscure.

Generally speaking, the quark propagator with external sources is
much more complex than that we considered here. Explicitly, it
should be Lorentz and gauge covariant\cite{an} and can be
decomposed in the form of the Lorentz structure, i.e., the
summation of the vector, axial-vector, scalar, pseudoscalar and
tensor terms. It is noticed that the conclusion is much more
complex and the physical meanings of the coefficients still
deserve to be investigated, we shall discuss it elsewhere.

Besides the anomalous action with pseudoscalar mesons, this method
can be extended to the processes with resonances such as the
vector mesons and axia-vector mesons are incorporated into the
chiral effective theory by the hidden local symmetry
method\cite{hidden,MWW-hidden}. The leading gauge invariant
anomalous action with external gauge fields was constructed
in\cite{C-W} by adopting the topological method with considering
the t'Hooft matching condition.

%%%%%%%%%%%%%%%%%%%%%%%%%%%%%%%%%%%%%%%%%%%%%%%%%%%%%%%%%%%%%%%%%%%%%%%%%%%%%%%%%%%%%%%%%%
\acknowledgments

\label{sec Ack}

This work was supported in part by the key projects of Chinese
Academy of Sciences(CAS), the National Science Foundation of China
(NSFC) and Specialized Research Fund for the Doctoral Program of
Higher Education of China.

%%%%%%%%%%%%%%%%%%%%%%%%%%%%%%%%%%%%%%%%%%%%%%%%%%%%%%%%%%%%%%%%%%%%%%%%%%%%%%%%%%%%%%%%%%%%


\begin{thebibliography}{999}
\bibitem{ma02} Yong-Liang Ma and Qing Wang, Phys.Lett.{\bf B},
188(2002).
\bibitem{folk-theorem} S.Weinberg, Physica {\bf A96}, 327(1979).
\bibitem{Gasser} J.Gasser, H.Leutwyler, Ann.Phys.(N.Y.){\bf 158}, 142(1984).\\
J.Gasser, H.Leutwyler, Nucl.Phys.{\bf B250}, 465(1985).
\bibitem{Gell-Mann} M.Gell-Mann, R.J.Oakes and B.Renner,
Phys.Rev.{\bf 175}, 2195(1968).
\bibitem{beforeQCD} S.Coleman, J.Wess and B.Zumino, Phys.Rev.{\bf
177}, 2239(1969).\\
C.G.Callan, S.Coleman, J.Wess and B.Zumino, Phys.Rev.{\bf 177},
2247(1969).\\
S.Weinberg, Phys.Rev.Lett.{\bf 18}, 188(1967).
\bibitem{WangQ1} Qing Wang, Yu-Ping Kuang, Ming Xiao and Xue-Lei Wang, Phys.Rev.{\bf
D61},054011(2000).
\bibitem{H.Yang} Hua Yang, Qing Wang, Yu-Ping Kuang and Qin Lu,
Phys. Rev.{\bf D66}, 014019(2002).
\bibitem{H.Yang1}
Hua Yang, Qing Wang and Qin Lu, Phys. Lett. {\bf B532}, 240(2002)
\bibitem{anom-norm} J.Balog, Phys.Lett.{\bf B149}, 197(1984).\\
A.A.Andrianov, L.Bonora, Nucl.Phys.{\bf B233}, 232(1985).\\
A.A.Andrianov, Phys.Lett.{\bf B157}, 425(1985).\\
A.A.Andrianov, et al.,Phys.Lett.{\bf B186}, 401(1987).\\
D.Espriu, E.De Rafael, J.Taron, Nucl.Phys.{\bf B345}, 22(1990).
\bibitem{wess} J.Wess and B.Zumino, Phys.Lett.{\bf B37}, 98(1971).
\bibitem{witten} E.Witten, Nucl.Phys.{\bf B223}, 442(1983).
\bibitem{topological} B. Zumino, UCB-PTH-83/16, LBL-16747, 1983; \\
K.C. Chou, H.Y. Guo, K. Wu, X.C. Song, Phys.Lett.{\bf B134},
67(1984); \\
L.Alvarez-Gaume and P.Ginsparg, Nucl.Phys.{\bf B 449},(1984).
\bibitem{C-W} K.C. Chou, Y.L. Wu and Y.B. Xie,  Mod. Phys. Lett. {\bf A1}
23(1986).
\bibitem{prove-p6} J.F.Donoghue, D.Wyler, Nucl.Phys.{\bf B316},
289(1989).\\
 J.Bijnens, A.Bramon, F.Cornet, Phys.Rev.Lett.{\bf 55},
1453(1988).
\bibitem{p6-form} J.Bijnens, A.Bramon, F.Cornet, Z.Phys.{\bf C46}, 599(1990).\\
R.Akhoury, A.Alfakih, Ann.Phys.(N.Y.){\bf 210}, 81(1991).
\bibitem{anom-constituent} R.D.Ball, Phys.Rep.{\bf 182},
1(1989).\\
J.Bijnens, Nucl.Phys.{\bf B367}, 709(1991).\\
J.Bijnens, Int.J.Mod.Phys.{\bf A8}, 3045(1993).\\
D.W.McKay, H.J.Munczek, Phys.Rev.{\bf D30}, 1825(1984).\\
D.W.McKay, H.J.Munczek, Phys.Rev.{\bf D32}, 266(1985).
\bibitem{DPT} H.Pagels and S.Stokar, Phys.Rev.{\bf D20},
2947(1979).
\bibitem{anom-review} J.Bijnens, Int.J.Mod.Phys.{\bf A8},
3045(1993).
\bibitem{pole1} D.Johnston,  LPTHE Orsay 86/49.
\bibitem{Landau} J.C.R.Bloch, Phys.Rev.{\bf D66}:034032,(2002).
\bibitem{axial} J.S.Ball, F.Zachariasen, Phys.Lett.{\bf
B106}:133,(1981).\\
J.R.Cudell, A.J.Gentles and D.A.Ross, Nucl.Phys.{\bf
B440}:521-542,(1995).
\bibitem{covariant} V.S.Gogokhia, Phys.Rev.{\bf D40}:4157,(1989), Phys.Rev.{\bf
D41}:3279,(1990).
\bibitem{lattice1} J.C.R.Bloch, Phys.Rev.{\bf D 66}, 034032(2002).
\bibitem{lattice2} Jon Ivar Skullerud, Anthony G.Williams, Phys.Rev.{\bf D63}, 054508(2001).
\bibitem{CSB} K.Lane, Phys.Rev.{\bf D10}, 2605(1974).\\
H.D.Politzer. Nucl.Phys.{B117}, 397(1976).\\
H.Pagels, Phys.Rev.{D19}, 3080(1979).\\
C.Callen, R.Dashen and D.Gross, Phys.Rev.{bf D17},
2717(1978).\\
D.G.Caldi, Phys. Rev.Lett.{\bf 39}, 121(1977).\\
R.D.Carlitz and D.B.Creamer, Ann.Phys.(N.Y.){\bf118}, 429(1979).
\bibitem{ball}R.D.Ball, Phys.Rep.{\bf 182}, 1(1989).
\bibitem{hidden}M.Bando, T.Kugo, S.Uehara, K.Yamawaki and T.Yanagida,
Phys.Rev.Lett.{\bf 54}: 1215 (1985).\\
%IS RHO MESON A DYNAMICAL GAUGE BOSON OF HIDDEN LOCAL SYMMETRY?%
M.Bando, T.Kugo and K.Yamawaki, Phys.Rept.{\bf
164}: 217 (1988). %NONLINEAR REALIZATION AND HIDDEN LOCAL SYMMETRIES%\\
\bibitem{MWW-hidden} Yong-Liang Ma, Qing Wang and Yue-Liang Wu, Eur.Phys.J.{\bf C39}:
201 (2005).
\bibitem{GCM} C.D.Roberts, R.T.Cahill and J.Praschifka, Annals Phys.{\bf
188}:20,(1988).
\bibitem{Bardeen} W.A.Bardeen, Phys.Rev.{\bf 184}, 1848(1969).
\bibitem{an} Y.H.An, H.Yang and Q. Wang, Eur.Phys.J.{\bf C29},
65(2003).
\bibitem{ESP} Qin Lu, Hua Yang and Qing Wang, hep-ph/0207128.
\bibitem{extra-dim}  O.Alvarez, Nucl.Phys.{\bf B238}: 61,(1984).

\bibitem{Fujikawa} K.Fujikawa, Phys.Rev.Lett.{\bf 42},
1195(1979).\\
K.Fujikawa, Phys.Rev.{\bf D21}, 2848(1980).
\bibitem{MW-chiral} Yong-Liang Ma and Yue-Liang Wu, hep-ph/0509083.
\bibitem{LR} Y.L.Wu, Int. J. Mod. Phys. A18 (2003) 5363-5419. \\
Y.L.Wu, Mod. Phys. Lett. A19 (2004) 2191.








\end{thebibliography}
\end {document}